\documentclass[letter,twocolumn]{jpsj3}

\usepackage{graphicx,color}
\usepackage{amsmath}
\usepackage{amssymb}
\usepackage{bm}
\usepackage{ulem}
\usepackage{comment}

\usepackage{txfonts}

\def\Vec#1{\bm{#1}}
\def\Hc2{H_\mathrm{c2}}
\def\SRO{\mathrm{Sr}_2\mathrm{RuO}_4}
\def\Tc{T_\mathrm{c}}

\begin{document}

\title{Searching for gap zeros in Sr$_2$RuO$_4$\\ via field-angle-dependent specific-heat measurement}

\author{
	Shunichiro \textsc{Kittaka}$^1$\thanks{kittaka@issp.u-tokyo.ac.jp}, 
	Shota \textsc{Nakamura}$^{1}$\thanks{Present address: Department of Engineering Physics, Electronics and Mechanics, Graduate School of Engineering, Nagoya Institute of Technology, Nagoya 466-8555, Japan}, 
	Toshiro \textsc{Sakakibara}$^{1}$, 
	Naoki \textsc{Kikugawa}$^{2}$, 
	Taichi \textsc{Terashima}$^{2}$, 
	Shinya \textsc{Uji}$^{2,3}$, 
	Dmitry A. \textsc{Sokolov}$^{4}$, 
	Andrew P. \textsc{Mackenzie}$^{4}$, 
	Koki \textsc{Irie}$^{5}$, 
	Yasumasa \textsc{Tsutsumi}$^{6,7}$, \\
	Katsuhiro \textsc{Suzuki}$^{8}$, and 
	Kazushige \textsc{Machida}$^{5}$ 
}

\inst{$^{1}$Institute for Solid State Physics, University of Tokyo, Kashiwa, Chiba 277-8581, Japan\\
      $^{2}$National Institute for Materials Science, 3-13 Sakura, Tsukuba, Ibaraki 305-0003, Japan\\
      $^{3}$Graduate School of Pure and Applied Sciences, University of Tsukuba, Tsukuba, Ibaraki 305-8577, Japan\\
      $^{4}$Max Planck Institute for Chemical Physics of Solids, Nothnitzer Str. 40, 01187 Dresden, Germany\\
      $^{5}$Department of Physics, Ritsumeikan University, Kusatsu, Shiga 525-8577, Japan\\
      $^{6}$Department of Basic Science, The University of Tokyo, Meguro, Tokyo 153-8902, Japan\\
      $^{7}$Condensed Matter Theory Laboratory, RIKEN, Wako, Saitama 351-0198, Japan\\
      $^{8}$Research Organization of Science and Technology, Ritsumeikan University, Kusatsu, Shiga 525-8577, Japan
}

\date{\today}

\abst{
The gap structure of $\SRO$, which is a longstanding candidate for a chiral $p$-wave superconductor, has been investigated 
from the perspective of the dependence of its specific heat on magnetic field angles at temperatures as low as 0.06~K ($\sim 0.04\Tc$).
Except near $\Hc2$, its fourfold specific-heat oscillation under an in-plane rotating magnetic field 
is unlikely to change its sign down to the lowest temperature of 0.06~K. 
This feature is qualitatively different from nodal quasiparticle excitations of a quasi-two-dimensional superconductor possessing vertical lines of gap minima.
The overall specific-heat behavior of $\SRO$ can be explained by 
Doppler-shifted quasiparticles around horizontal line nodes on the Fermi surface, 
whose in-plane Fermi velocity is highly anisotropic, 
along with the occurrence of the Pauli-paramagnetic effect.
These findings, in particular, the presence of horizontal line nodes in the gap, call for a reconsideration of the order parameter of $\SRO$.
}

\maketitle


$\SRO$, a layered-perovskite superconductor with $\Tc$=1.5~K,\cite{Maeno1994Nature} has attracted enormous attention ever since 
Knight-shift experiments provided favorable evidence that it exhibits spin-triplet pairing.\cite{Ishida1998Nature,Duffy2000PRL,Ishida2008JPCS,Ishida2015PRB} 
Numerous experiments have demonstrated that $\SRO$ has non $s$-wave properties,\cite{Mackenzie2003RMP,Maeno2012JPSJ} and 
some of the experimental reports indicate a degenerate order parameter.\cite{Luke1998Nature,Xia2006PRL}
The simple Fermi-surface topology of $\SRO$ comprising of three cylindrical sheets ($\alpha$, $\beta$, and $\gamma$)~\cite{Bergemann2003AP,Damascelli2000PRL} 
together with its well-characterized Fermi-liquid behavior
has led to the construction of several theoretical models to describe superconductivity.\cite{Mackenzie2003RMP}
Among these models, a spin-triplet chiral $p$-wave pairing characterized by $\Vec{d}=\Delta_0\hat{\Vec{z}}(k_x \pm ik_y)$ has been considered to be a promising candidate.

However, several experimental facts exist that cannot be explained in the framework of this spin-triplet scenario.\cite{Mackenzie2017QM}
A serious controversy is the mechanism of the first-order superconducting transition along with the $\Hc2$ limit induced by the in-plane magnetic field.\cite{Yonezawa2013PRL,Yonezawa2014JPSJ,Kittaka2014PRB,Kikugawa2016PRB}
It is reminiscent of the Pauli-paramagnetic effect that is not allowed in the spin-triplet scenario. 
The superconducting gap structure of $\SRO$ is also a contentious issue.
In general, a chiral $p$-wave gap opening on the cylindrical Fermi-surface sheets has no symmetry-protected node.
Nevertheless, the gap amplitude of $\SRO$ has been widely accepted to be modulated, and lines of deep minima (or nodes) are suggested to be present somewhere in the gap 
because of the power-law temperature dependence of various physical quantities.\cite{Nishizaki2000JPSJ,Suzuki2002PRL,Bonalde2000PRL,Ishida2000PRL}
Furthermore, universal heat transport has raised the possibility of a nodal gap.\cite{Suzuki2002PRL}
Various gap structures including vertical and horizontal line node gaps have been proposed so far;\cite{Izawa2001PRL,Tanatar2001PRL,Deguchi2004PRL,Deguchi2004JPSJ,Hassenger2017PRX}
however, the location of gap minima has not yet been established.

During this decade, field-angle-dependent measurements that probe quasiparticle density of states, $N(E)$, have been developed as powerful tools for determining the position of gap zeros.\cite{Sakakibara2016RPP,Matsuda2006JPCM,Vorontsov2006PRL,Hiragi2010JPSJ}
This technique takes advantage of the Doppler energy shift, $\delta E = m_{\rm e}\Vec{v}_{\rm F} \cdot \Vec{v}_{\rm s}$, of $N(E)$ in the mixed state.
Here, $m_{\rm e}$ is the effective mass, $\Vec{v}_{\rm F}$ is the Fermi velocity, and $\Vec{v}_{\rm s}$ ($\perp \Vec{H}$) is the velocity of the supercurrent circulating around vortices.
If the gap has zeros somewhere on the Fermi surface, $N(E=0)$ becomes finite in those areas because of the Doppler shift and
varies with the angle between the field ($\perp \Vec{v}_{\rm s}$) and the nodal ($\Vec{v}_{\rm F}$) directions.
Therefore, the field-angle dependence of $N(E=0)$ provides key information about the gap structure.

In 2004, Deguchi \textit{et al}. reported the in-plane field-angle $\phi$ dependence of the specific heat, $C(\phi)$, of $\SRO$ in the temperature range $0.12 \le T \le 0.51$~K.\cite{Deguchi2004PRL,Deguchi2004JPSJ}
They proposed that the gap has fourfold anisotropy within the $ab$ plane, \textit{i.e.}, four vertical lines of gap minima, 
based on the assumption that $C(\phi) \propto N(\phi,E=0)$.
However, recent theoretical studies~\cite{Vorontsov2006PRL,Hiragi2010JPSJ} have suggested that
the Doppler-predominant condition [$C \propto N(E=0)$] is restricted in the low-temperature region because
$C(T)$ at a finite temperature mainly reflects $N(E)$ at $E \sim 2.4k_{\rm B}T$.
In other words, the reversed $\phi$ dependence of $N(E\ne 0)$ changes the sign of the $C(\phi)$ oscillation, 
\textit{e.g.}, approximately at $0.1\Tc$ in the case of quasi-two-dimensional $d_{x^2-y^2}$-wave superconductors.\cite{Vorontsov2006PRL,Hiragi2010JPSJ}
Such a sign change was indeed observed in nodal superconductors CeCoIn$_5$~\cite{An2010PRL} and KFe$_2$As$_2$,\cite{Kittaka2014JPSJ}
though it has not yet been detected in $\SRO$, at least above 0.12~K ($\sim 0.08\Tc$).\cite{Deguchi2004PRL,Deguchi2004JPSJ}

Here, we report the results of high-precision $C(\phi)$ measurements at temperatures as low as 0.06 K ($\sim 0.04\Tc$).
We reveal that, well below $\Hc2$, the normalized amplitude of the $C(\phi)$ oscillation, $A_4(T,H)$, monotonically varies with $T$ and $H$ without a sign change. 
By comparing our results with microscopic calculations, 
we find that the observed features in $C(T,H,\phi)$  can be qualitatively reproduced 
by a horizontal line node gap on a Fermi-surface sheet, which exhibits a strong in-plane anisotropy in the Fermi velocity, along with the Pauli-paramagnetic effect.

\begin{figure}
\begin{center}
\includegraphics[width=3.4in]{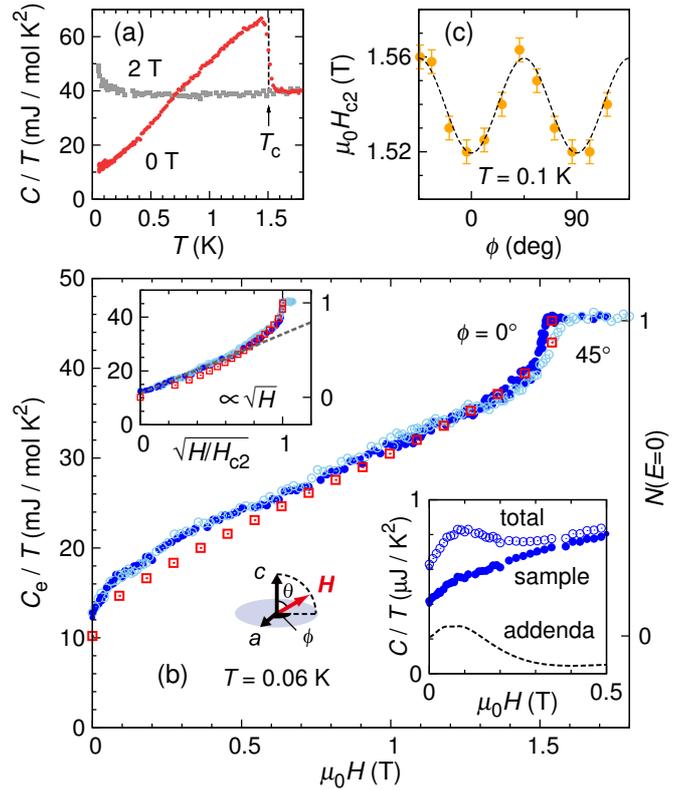}
\end{center}
\vspace{-0.2in}
\caption{
(a) Temperature dependence of the specific heat divided by temperature $C/T$ at 0 and 2~T in $H \parallel [100]$. 
(b) $C_{\rm e}/T=(C-C_{\rm ph}-C_{\rm N})/T$ at 0.06~K as a function of the magnetic field in $H \parallel [100]$ ($\phi=0^\circ$; closed circles) and $H \parallel [110]$ ($\phi=45^\circ$; open circles).
Squares represent $N(E=0)$ at $\phi=45^\circ$ obtained by microscopic calculation assuming $\Delta(\Vec{k})=\Delta_0\cos ck_z$ with $b=0.5$ and $\mu=0.04$ (refer to the text).
The upper inset shows the same data as a function of $\sqrt{H/\Hc2}$ along with a linear function in $\sqrt{H}$ (dashed line).
The lower inset shows the field dependence of the addenda heat capacity (dashed line) together with 
total (open circles) and sample (closed circles) heat capacities (Sr$_2$RuO$_4$; at 0.06 K in $H \parallel [100]$).
(c) In-plane upper critical field $\Hc2$ as a function of the azimuthal field angle $\phi$ at 0.1~K.}
\label{samp}
\end{figure}

High-quality single crystals of $\SRO$ were grown by a floating-zone method,\cite{Mao2000MRB} and a single 11.2-mg piece was used.
The crystal was oriented using the backscattering X-ray Laue method.
The angle-resolved specific heat $C(T,H,\phi,\theta)$ was measured by the quasi-adiabatic heat-pulse method using a dilution refrigerator.
Here, $\phi$ ($\theta$) denotes an azimuthal (polar) field angle relative to the $[100]$ ($[001]$) axis, as represented in Fig.~\ref{samp}(b).
The addenda of our calorimeter mainly consisted of a stage (silver foil), thermometer, and heater.
A ruthenium-oxide chip resistor (Panasonic, ERJ-XGNJ, 8.2~k$\Omega$) was used as a thermometer; 
it was cut into thirds (resulting in reducing the resistance to 3.3~k$\Omega$) and 
whose substrate was polished to reduce its heat capacity.
A ruthenium-oxide chip resistor (ROHM, MCR004, 240~$\Omega$) was used as a heater, whose substrate was also polished.
In this study, we have measured the addenda heat capacity carefully and subtracted its contribution,
as depicted in the lower inset of Fig.~\ref{samp}(b). 
The magnetic field was generated using a vector magnet; the field was up to 3~T along the $z$-axis and 5~T along the $x$-axis. 
By using a stepper motor mounting on top of the Dewar, the refrigerator could be rotated around the $z$-axis.
Thus, the orientation of the magnetic field was controlled three-dimensionally with high accuracy (better than $0.05^\circ$).

Figure \ref{samp}(a) shows the temperature dependence of $C/T$ at 0 and 2~T in $H \parallel [100]$.
In zero field, a clear specific-heat jump is observed at $\Tc$ of 1.505~K (midpoint), which is as high as the highest $\Tc$ of $\SRO$.\cite{Mackenzie1998PRL}
At 2~T ($H > \Hc2$) and at low temperatures, $C/T$ shows a gradual increase with decreasing temperature.
This can be attributed to the nuclear specific heat.
To explore the electronic contribution $C_{\rm e}$, 
the phonon and nuclear contributions, $C_{\rm ph}$ and $C_{\rm N}$, respectively, are subtracted from the data shown below. 
Here, the Debye temperature is 410~K, and $C_{\rm N}$ is $(0.08+0.14H^2)/T^2$~$\muup$J/(mol K).
The latter is calculated using a nuclear spin Hamiltonian with the nuclear quadrupole resonance parameters of $^{99}$Ru, $^{101}$Ru, and $^{87}$Sr nuclei.\cite{Ishida2001PRB,Murakawa2004PRL}
The resulting $C_{\rm e}/T=(C-C_{\rm ph}-C_{\rm N})/T$ at 2~T still shows a slight upturn at low temperatures (not shown),
implying insufficient subtraction of the non-electronic contributions.
Nevertheless, we adopt this definition to avoid uncertainty.

In Fig.~\ref{samp}(b), $C_{\rm e}/T$ at 0.06~K is plotted as a function of $H$ applied parallel to the $[100]$ and $[110]$ axes.
The increase in $C_{\rm e}(H)/T$ from 0~T to $\Hc2$ is 33~mJ/(mol K$^2$),
which is comparable to results of a previous study,\cite{Deguchi2004PRL}
although the absolute value is enhanced due to the difference in the definition of $C_{\rm e}$.\cite{Kittaka2017c}
This fact indicates that the superconducting volume fraction of the present sample is above 90\%, 
and that an offset of $\sim 6$~mJ/(mol K$^2$) is given to $C_{\rm e}/T$ in Fig. 1(b) by the remaining non-electronic contributions.
In the high-field region, $C_{\rm e}(H)$ shows a single, sharp jump at $\mu_0\Hc2$ of 1.52 (1.56)~T in $H \parallel [100]$ ($[110]$). 
In Fig.~\ref{samp}(c), the in-plane $\Hc2$ shows fourfold anisotropy, which is consistent with the results of previous reports.\cite{Mao2000PRL,Yaguchi2002PRB,Yonezawa2014JPSJ}
At low fields, $C_{\rm e}(H)$ is nearly proportional to $\sqrt{H}$, as indicated with a dashed line in the upper inset of Fig.~\ref{samp}(b).
This fine $\sqrt{H}$ behavior supports the occurrence of low-energy quasiparticle excitations around lines of deep gap minima or nodes.\cite{Volovik1993JETPL}

In previous reports,\cite{Nishizaki2000JPSJ,Deguchi2004PRL} a shoulder anomaly was detected in the low-temperature $C_{\rm e}(H)$ around $\mu_0H \sim 0.15$~T,
which was attributed to the abrupt suppression of minor gaps in a weakly-coupled multiband superconductor. 
This is similar to the cases of MgB$_2$~\cite{Bauquet2002PRL} and KFe$_2$As$_2$.\cite{Kittaka2014JPSJ, Hardy2014JPSJ}
However, no such anomaly is observed in our data after precise subtraction of the addenda contribution.
It is noted that the total heat capacity shows a shoulder anomaly which can be attributed to the addenda contribution [the lower inset of Fig.~\ref{samp}(b)].
The lack of the multigap feature indicates relatively strong coupling between the three gaps on the $\alpha$, $\beta$, and $\gamma$ bands;
all three gaps survive up to relatively high fields.

\begin{figure}[t]
\begin{center}
\includegraphics[width=3.4in]{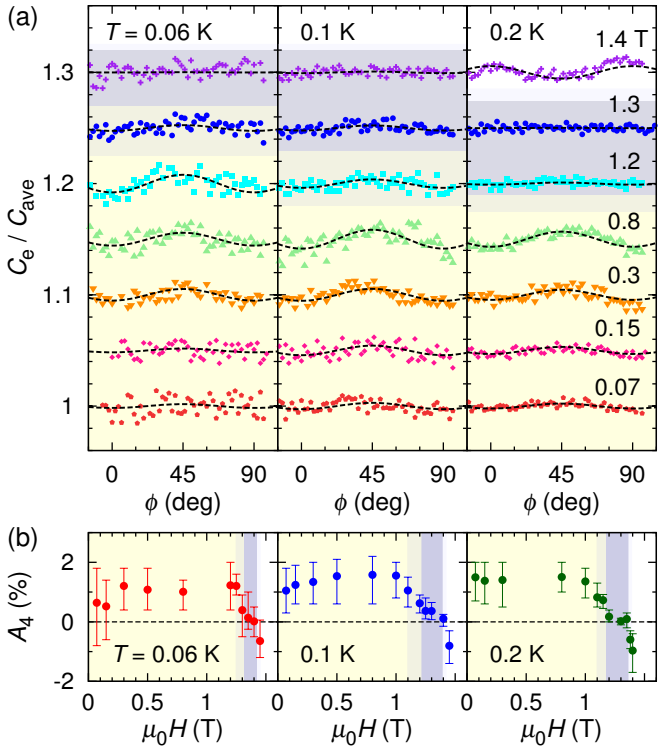}
\end{center}
\vspace{-0.2in}
\caption{
Field-angle $\phi$ dependence of $C_{\rm e}/C_{\rm ave}$ at 0.06, 0.1, and 0.2~K measured in a rotating magnetic field within the $ab$ plane. 
Here, $C_{\rm ave}=C_0+C_H$ and
the data are vertically shifted by 0.05 for clarity.
Dashed lines are fits to the data based on the function $C=C_0+C_H(1-A_4\cos4\phi)$. 
Here, $C_0$ and $C_H$ are zero-field and field-dependent components of $C_{\rm e}$.
(b) $A_4(H)$ obtained from the fits.
}
\label{phi}
\end{figure}

To examine the in-plane gap anisotropy, 
$C_{\rm e}(\phi)$ was measured in several magnetic fields ($0.07 \le \mu_0H \le 1.45$~T) that were rotated within the $ab$ plane.
The results of $C_{\rm e}(\phi)/T$ at 0.06, 0.1, and 0.2~K are shown in Fig.~\ref{phi}(a).
In the wide field range, $C_{\rm e}(\phi)$ shows a clear fourfold oscillation that fits the function $C_{\rm e}(\phi)=C_0+C_H(1-A_4\cos4\phi)$ [dashed lines in Fig.~\ref{phi}(a)].\cite{Kittaka2018a} 
Here, $C_0$ is the zero-field value of $C_{\rm e}$, $C_H$ is the field-dependent part of $C_{\rm e}$, 
and $A_4$ is the amplitude of the fourfold oscillation normalized by $C_H$.
In Fig.~\ref{phi}(b), $A_4(H)$ obtained from the fit in the entire range is plotted at each temperature
along with somewhat overestimated error bars; 
the error bars represent minimum and maximum values of $A_4(H)$ obtained from the partial fit in the range $\phi_0-30^\circ \le \phi\le \phi_0+30^\circ$,
where the central angle $\phi_0$ is changed from $30^\circ$ to $60^\circ$ approximately.
In the present temperature range, $A_4(H)$ stays roughly 1\% below 1~T (yellow shaded area).
Importantly, a sign change of $A_4$ is  unlikely to occur at low fields;
\textit{e.g.}, $A_4$ including the error bars is always positive for 0.06~K $\le T \le$ 0.2~K at a low field of 0.3~T. 
With increasing field, $A_4(H)$ suddenly decreases and becomes nearly zero around 1.3~T (blue shaded area).
In high fields close to $\Hc2$, the sign of $A_4(H)$ is reversed due to the anisotropy of $\Hc2(\phi)$.

In order to explore the out-of-plane gap anisotropy, 
we investigated the polar-angle $\theta$ dependence of the specific heat at 0.06~K under a rotating field within the (010) plane ($\phi=0^\circ$).
As demonstrated in Supplemental Material~\cite{KittakaSM} (I), however,
little information on the out-of-plane gap anisotropy can be extracted from $C_{\rm e}(\theta)$~\cite{Kittaka2017b} 
because $C_{\rm e}(\theta,H)$ is dominated by $H_{\parallel c}$ because of the large anisotropy ratio of the coherence length ($\xi_a/\xi_c \sim 60$).\cite{Kittaka2014PRB}
This fact, however, suggests that 
the previously reported steep suppression of $A_4(\theta)$ under a conically-rotating field~\cite{Deguchi2004JPSJ} is 
not due to the compensation of antiphase gap anisotropies between active and passive bands but due to this scaling by $H_{\parallel c}$.

\begin{figure}[t]
\begin{center}
\includegraphics[width=3.4in]{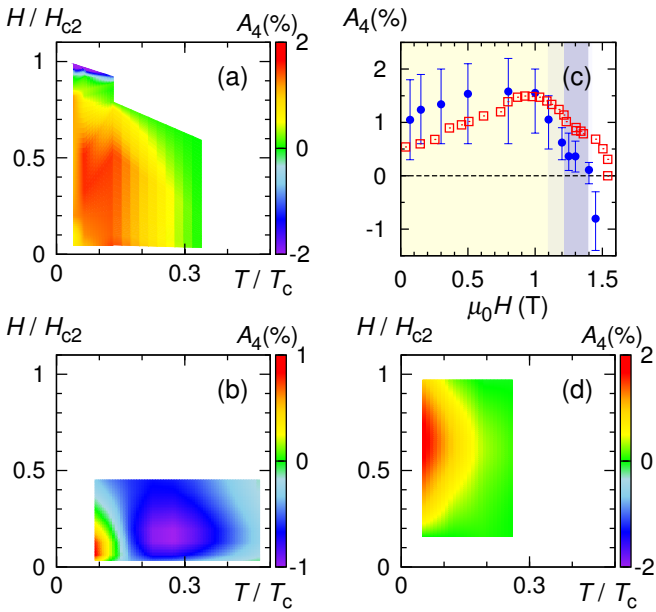}
\end{center}
\vspace{-0.2in}
\caption{
(a) Contour plot of $A_4(T,H)$ made by using the present ($T/\Tc \le 0.13$) and previous ($T/\Tc = 0.34$, $H/\Hc2 \le 0.6$)~\cite{Deguchi2004PRL} experimental data. 
(b) $A_4(T,H)$ calculated for a $d_{xy}$-wave gap with $b=0$ and $\mu=0$.\cite{Hiragi2010JPSJ}
$A_4(T,H)$ for a $d_{x^2-y^2}$-wave gap is obtained by reversing the sign in~(b).
(c) $A_4(H)$ (squares) and (d) $A_4(T,H)$ map obtained from microscopic calculations assuming $\Delta(\Vec{k})=\Delta_0\cos ck_z$ with $b=0.5$ and $\mu=0.04$~\cite{Machida2017,Machida2017b} (see text).
Circles in (c) are the experimental data at 0.1~K.
}
\label{A4}
\end{figure}

Let us discuss the origin of the $C_{\rm e}(\phi)$ oscillation in $\SRO$.
By combining the present $A_4(H)$ data at 0.06, 0.1, and 0.2~K with $A_4=0$ at 0.51~K below 0.9~T,\cite{Deguchi2004PRL}
a contour plot of $A_4(T,H)$ is depicted in Fig.~\ref{A4}(a).
Clearly, this map is very different from $A_4(T,H)$ calculated for $d_{xy}$- and $d_{x^2-y^2}$-wave gaps on a rippled cylindrical Fermi surface [Fig.~\ref{A4}(b)].\cite{Hiragi2010JPSJ}
In particular, Fig.~\ref{A4}(b) shows a sign change of $A_4$ at $T \sim 0.12\Tc$ in low fields below $\sim 0.3\Hc2$.
On theoretical grounds, the sign-changing line in Fig.~\ref{A4}(b) moves by warping the cylindrical Fermi-surface shape;\cite{Vekhter2008PhysicaB}
the line is shifted toward a lower (higher) field and a lower (higher) temperature if vertical line nodes are present on large (small) curvature parts of the warped Fermi surface.\cite{Machida2017}
For $\SRO$, the quasi-two-dimensional $\gamma$ band is nearly cylindrical, which yields the sign-changing line around approximately $0.1\Tc$. 
By contrast, quasi-one-dimensional $\alpha$ and $\beta$ bands have flat and high-curvature parts in the $\langle 100 \rangle$ and $\langle 110 \rangle$ direction, respectively.
If the $d_{xy}$-type vertical lines of deep gap minima exist on the $\alpha$ and/or $\beta$ bands and dominantly contribute to the $C_{\rm e}(\phi)$ oscillation, 
the observed $A_4(T,H)$ map might be reproduced to some extent.
However, this is not guaranteed because fine-tuning of multiband parameters is essential.

An alternative, more promising scenario is to assume a horizontal line node gap.
According to the recent band-structure calculation,\cite{Steppke2017Science} 
the in-plane $v_{\rm F}$-anisotropy ratio, $v_{{\rm F} \parallel [110]}/v_{{\rm F} \parallel [100]}$, 
is estimated to be roughly 4 (0.8~\cite{Kittaka2017d}) for the $\gamma$ band ($\alpha$ and $\beta$ bands).
If the in-plane anisotropy of the Fermi velocity, $v_{\rm F}(\phi_k)$, is sufficiently large,
a prominent $C_{\rm e}(\phi)$ oscillation is expected because of an anisotropic Doppler shift $|\delta E(\phi_k)| \propto v_{\rm F}(\phi_k) |\sin(\phi - \phi_k)|$ at $\phi_k$ around the horizontal line nodes.\cite{Vekhter1999PRB}
To examine this possibility, microscopic calculations were performed
by assuming a horizontal line node gap $\Delta(\Vec{k})=\Delta_0\cos ck_z$ and $v_{\rm F}(\phi_k)=v_{\rm F0}(b)(1 - b\cos4\phi_k)$~\cite{Machida2017} on the rippled cylindrical Fermi surface; 
this method was similar to that used in previous reports.\cite{Miranovic2005JPC,Hiragi2010JPSJ,Amano2014PRB}
Here, we use the anisotropic parameter $b=0.5$ (for the $\gamma$ band) and the Maki parameter $\mu=0.04$;
the latter is adopted to reproduce the $\Hc2$ limit.

The calculated result of $N(E=0,H)$ in $H \parallel [110]$ is shown in Fig.~\ref{samp}(b) with squares that match $C_{\rm e}(H)$ sufficiently 
owing to the nodal gap with the finite Maki parameter.\cite{Ichioka2007PRB,Machida2008PRB}
Slight mismatch at low fields can be attributed to thermal excitations of quasiparticles in $C_{\rm e}/T$ at 0.06 K which are absent in $N(E=0)$;
note that the origin of $N(E=0)$ is vertically shifted 
so that in zero field $N(E=0)$ corresponds to $C_{\rm e}/T$ at 0~K estimated by the extrapolation of the $C_{\rm e}(T)$ data to 0~K.
In Fig.~\ref{A4}(c), the calculated $A_4(H)$ (squares)~\cite{Machida2017,Machida2017b} is compared with the experimental result at 0.1~K (circles). 
The absence of a sign change in $A_4(H)$ at low fields and its steep decrease in high fields are captured by this model.  
The latter is caused by the Pauli-paramagnetic effect; without this, $A_4(H)$ starts to decrease at a relatively low field ($\sim\Hc2/3$) [refer to Supplemental Material \cite{KittakaSM} (II)]
and the low-temperature $C_{\rm e}(H)$ behavior as well as the $\Hc2$ limitation cannot be reproduced.
A slight mismatch above $\sim 1.2$~T might suggest the occurrence of an extra phase that is not considered in the present calculation.
A possible one is the Fulde-Ferrell-Larkin-Ovchinnikov phase.\cite{Suzuki2011JPSJ}
This phase might be the origin of the strange $A_4(H) \sim 0$ region around 1.3~T and in-plane $\Hc2$ anisotropy ($H_{\parallel [100]} < H_{\parallel [110]}$) developing below 0.8~K (above 1.2~T),\cite{Mao2000PRL,Kittaka2009PRB}
which is inverted to the anisotropy expected from $v_{\rm F}(\phi_k)$ of the $\gamma$ band.
Figure~\ref{A4}(d) shows a contour map of the calculated $A_4(T,H)$.
This also reproduces no sign change and peak structure of $A_4(T,H)$ in Fig.~\ref{A4}(a).
Thus, a horizontal line node gap can explain various features of $C_{\rm e}(T,H,\phi)$ 
by taking into account the in-plane $v_{\rm F}$ anisotropy of the $\gamma$ band~\cite{Kittaka2017e} and the Pauli-paramagnetic effect.

The horizontal line node gap is, however, incompatible with the $p$- and $f$-wave scenarios suggested in the previous reports.\cite{Deguchi2004PRL,Hassenger2017PRX}
In addition, the absence of multigap feature in $C_{\rm e}(H)$, 
together with the good match between the experimental $C_{\rm e}(H)$ and the calculated $N(E=0)$ assuming a nodal gap on a single band, 
supports the presence of nodes in all three gaps, rather than only in passive bands.~\cite{Zhitomirsky2001PRL}
In the tetragonal $D_{4h}$ symmetry, several gap symmetries possessing horizontal line nodes exist, 
\textit{e.g.}, $(5k_z^2-1)(k_x \pm ik_y)$ [$E_u$], $k_z(k_x \pm ik_y)$ [$E_g$], $k_z$ [$A_u$], and $\cos (k_zc)$ [$A_{1g}$] in weak spin-orbit coupling (the orbital part only).
The former two are broken time-reversal symmetries that are consistent with the results of $\muup$SR and Kerr-rotation experiments.\cite{Luke1998Nature,Xia2006PRL}
Non-zero residual thermal conductivity along the $c$ axis~\cite{Hassenger2017PRX} might support the $E_u$ or $A_{1g}$ symmetry,
for which horizontal line nodes are located at off-centered positions.
To test these gap symmetries, a spin resonance experiment using inelastic neutron scattering at $\Vec{Q}=(1/3,1/3,q_z)$ might be useful, \textit{e.g.}, at $q_z=0.15, 0.25$ and 0.35.
Notably, the absence of spin resonance at $\Vec{Q}=(1/3,1/3,0)$~\cite{Kunkemoller2017PRL} is also incompatible with the vertical-line-node scenario.

In summary, we have investigated the gap anisotropy of $\SRO$ from field-angle-dependent specific-heat measurements.
In $H \parallel ab$, the low-field specific heat increases in proportion to $\sqrt{H}$ with no multigap structure at 0.06~K.
This suggests that multiple gaps on the $\alpha$, $\beta$, and $\gamma$ bands equivalently survive and possess lines of deep minima in some directions.
We have revealed that the fourfold oscillation in the low-field $C_{\rm e}(\phi)$ does not change its sign even at 0.06~K.
In addition, an abrupt change in the normalized oscillation amplitude $A_4(H)$ is observed around $\mu_0H \sim 1.2$~T.
These features can be understood by considering a horizontal line node gap along with large in-plane anisotropy in $v_{\rm F}$ and the Pauli-paramagnetic effect.
The present results shed fresh light on a possibility that the gap symmetry possesses horizontal line nodes and 
challenge the current view of the order parameter of $\SRO$.\\ \\

\acknowledgments
We thank Y. Yoshida and H. Yaguchi for their support with the experiments. We also thank Y. Maeno and S. Yonezawa for the fruitful discussion.
A part of the numerical calculations was performed by using the HOKUSAI supercomputer system in RIKEN.
This work was supported by a Grant-in-Aid for Scientific Research on Innovative Areas ``J-Physics'' (15H05883,18H04306) from MEXT
and KAKENHI (18H01161, 15K05158, 15H03682, 26400360, 15K17715, 15J05698, 17K05553, 18K04715) from JSPS.

\clearpage
\onecolumn
\renewcommand{\thefigure}{S\arabic{figure}}
\setcounter{figure}{0}

\small
\begin{center}
{\normalsize Supplemental Material for \\
\textbf{Searching for gap zeros in Sr$_2$RuO$_4$ via field-angle-dependent specific-heat measurement}}\\
\vspace{0.1in}
Shunichiro Kittaka,$^{1}$ Shota Nakamura,$^{1}$ Toshiro Sakakibara,$^{1}$ Naoki Kikugawa,$^2$ Taichi Terashima,$^2$ Shinya Uji,$^{2,3}$ \\Dmitry A. Sokolov,$^4$ Andrew P. Mackenzie,$^4$ 
Koki Irie,$^5$ Yasumasa Tsutsumi,$^{6,7}$ Katsuhiro Suzuki,$^8$ and Kazushige Machida$^5$\\
{\small 
\textit{$^1$Institute for Solid State Physics, University of Tokyo, Kashiwa, Chiba 277-8581, Japan}\\
\textit{$^2$National Institute for Materials Science, 3-13 Sakura, Tsukuba, Ibaraki 305-0003, Japan}\\
\textit{$^3$Graduate School of Pure and Applied Sciences, University of Tsukuba, Tsukuba, Ibaraki 305-8577, Japan}\\
\textit{$^4$Max Planck Institute for Chemical Physics of Solids, Nothnitzer Str. 40, 01187 Dresden, Germany}\\
\textit{$^5$Department of Physics, Ritsumeikan University, Kusatsu, Shiga 525-8577, Japan}\\
\textit{$^6$Department of Basic Science, The University of Tokyo, Meguro, Tokyo 153-8902, Japan}\\
\textit{$^7$Condensed Matter Theory Laboratory, RIKEN, Wako, Saitama 351-0198, Japan}\\
\textit{$^8$Research Organization of Science and Technology, Ritsumeikan University, Kusatsu, Shiga 525-8577, Japan}\\
}
\end{center}

\section{Out-of-plane field-angle dependence of the specific heat for Sr$_2$RuO$_4$}

\begin{figure}
\begin{center}
\includegraphics[width=4.3in]{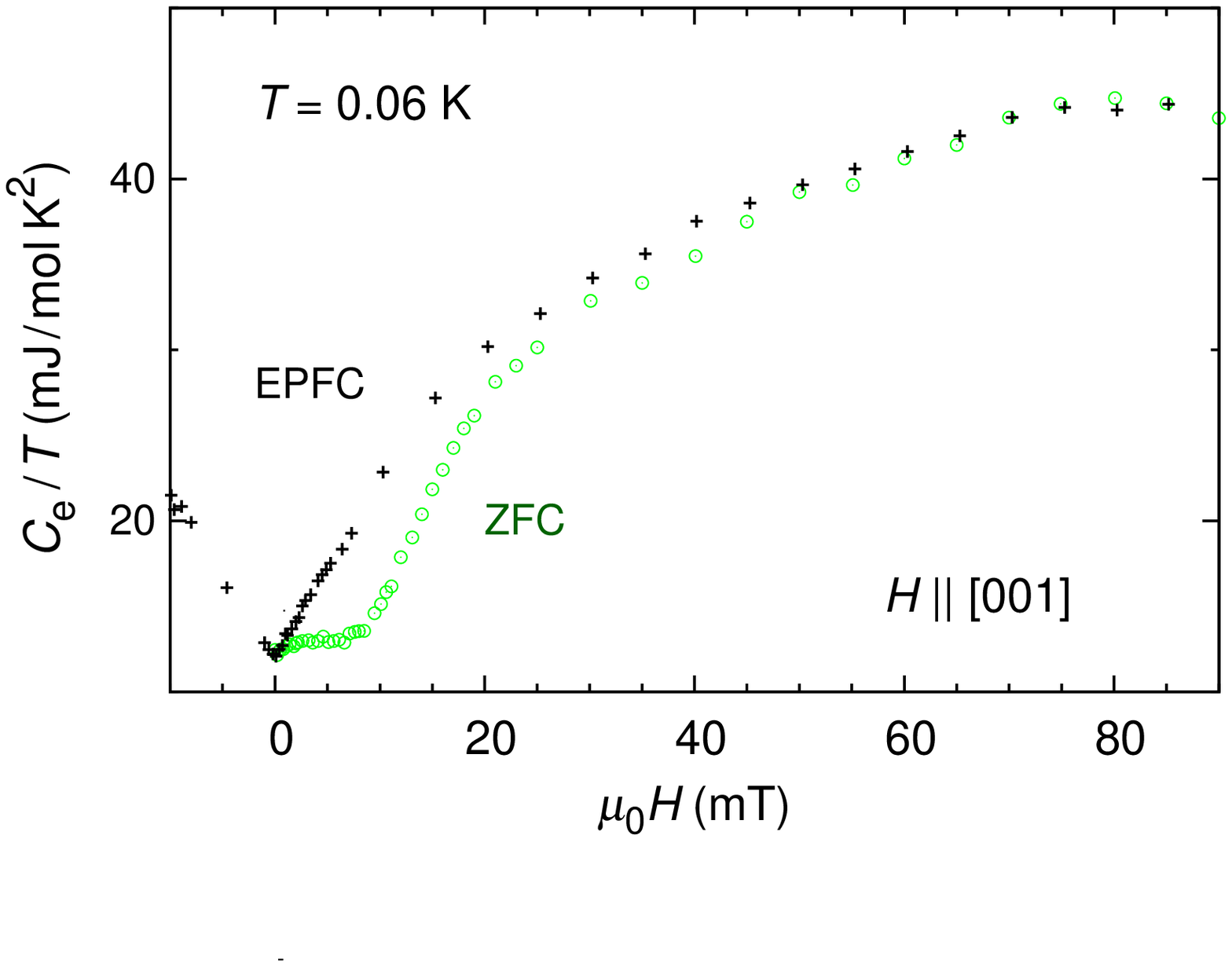} 
\caption{
$C_{\rm e}(H)/T$ at 0.06 K in $H \parallel [001]$ measured in the zero-field-cooling (ZFC; circles) and each-point-field-cooling (EPFC; crosses) processes.
}
\label{CH001}
\end{center}
\end{figure}

Figure~\ref{CH001} shows $C_{\rm e}(H)/T$ of Sr$_2$RuO$_4$ measured at 0.06~K in $H \parallel [001]$.
At low fields, $C_{\rm e}(H)$ measured after the zero-field-cooling process is nearly unchanged up to $\mu_0H_{\rm c1}$ ($\sim 7$~mT for $H \parallel [001]$) and rapidly increases above $H_{\rm c1}$.
By contrast, $C_{\rm e}(H)$ measured in the each-point-field-cooling (EPFC) process$^{1)}$ varies proportionally from 0~T to $H$.
This $H$-linear behavior is probably due to an increase of trapped fluxes inside the sample. 
These effects unfortunately mask low-energy quasiparticle excitations in $H \parallel [001]$, in sharp contrast to $C_{\rm e}(H)$ in $H \parallel ab$;
it is expected that $C_{\rm e}(H)$ increases in proportion to $\sqrt{H}$ in $H \parallel [001]$ if the gap has line nodes.

\begin{figure}
\begin{center}
\includegraphics[width=4.5in]{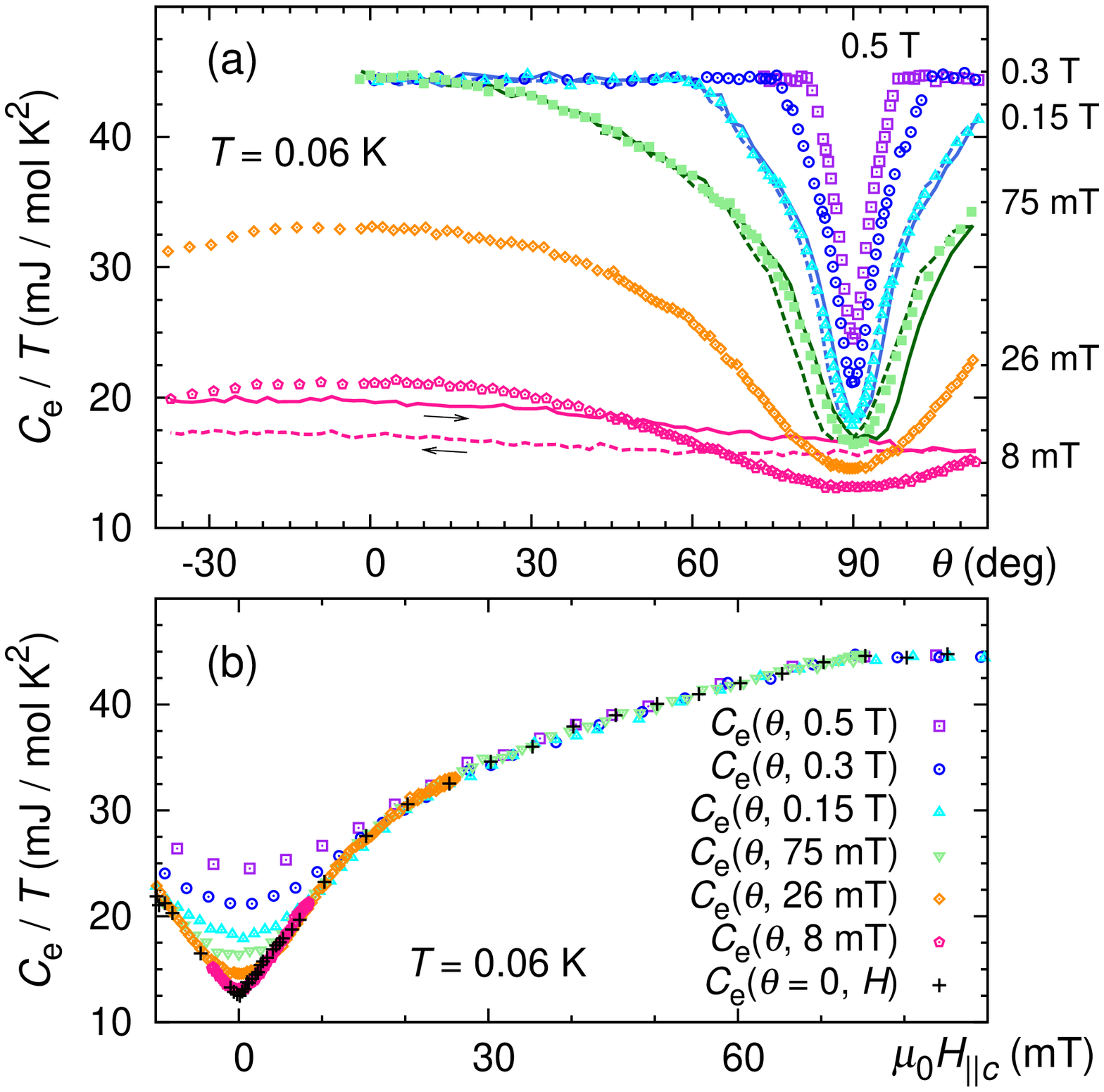} 
\caption{
(a) Field-angle $\theta$ dependence of $C_{\rm e}/T$ at 0.06~K in a rotating magnetic field within the $(010)$ plane.
Here, $\theta$ denotes the field angle with respect to the $[001]$ axis.
Symbols are the data taken in the EPFC process.
Solid and dashed lines are the data (8, 75, and 150 mT) taken in the increasing and decreasing $\theta$ sweeps, respectively, 
without increasing the sample temperature significantly.
(b) The same data (symbols), as in (a), plotted as a function of the magnetic-field component along the $[001]$ axis, $H_{\parallel c}$ ($=H\cos\theta$).
The EPFC $C_{\rm e}(H)/T$ data in $H \parallel [001]$ ($\theta=0$; crosses) are also plotted for comparison.
}
\label{theta}
\end{center}
\end{figure}

To explore the out-of-plane gap anisotropy, 
we investigated the polar-angle $\theta$ dependence of the specific heat at 0.06~K under a rotating field within the (010) plane ($\phi=0^\circ$).
First, we measured $C_{\rm e}(\theta)$ while keeping the sample at a temperature well below 1.5~K (\textit{i.e.}, not in the EPFC process).
Then, we found that the results strongly depend on the rotational direction of $H$;
this is indicated in Fig.~\ref{theta}(a) by solid and dashed lines 
that represent the data taken in increasing and decreasing $\theta$ sweeps, respectively. 
The dependence on the rotational direction becomes prominent at lower fields.
This can be attributed to the strong trap of the magnetic flux, which is illustrated in Fig.~\ref{CH001}.
Therefore, to align the trapped fluxes with the applied field orientation,
we then performed EPFC measurements of $C_{\rm e}(\theta)$; \textit{i.e.}, the sample temperature once increases well above $\Tc$ after each field rotation.
EPFC data of $C_{\rm e}(\theta)$ [symbols in Fig.~\ref{theta}(a)] become symmetric around $\theta=90^\circ$ with no rotational-direction dependence.

In Fig.~\ref{theta}(b), the EPFC data of $C_{\rm e}(\theta)$ at various fields are plotted as a function of $H_{\parallel c}=H\cos\theta$, 
together with the EPFC data of $C_{\rm e}(\theta=0,H)$.
Except near $H_{\parallel c} \sim 0$, 
all $C_{\rm e}(\theta,H)$ data are scaled by $H_{\parallel c}$, and the scaling function matches $C_{\rm e}(\theta=0,H)$ well.
This means that it is difficult to investigate the out-of-plane gap anisotropy of Sr$_2$RuO$_4$ from $C_{\rm e}(\theta)$ measurements because of the large anisotropy of the coherence length.

\section{Pauli-paramagnetic effect on $A_4(H)$}
\begin{figure}
\begin{center}
\includegraphics[width=4.5in]{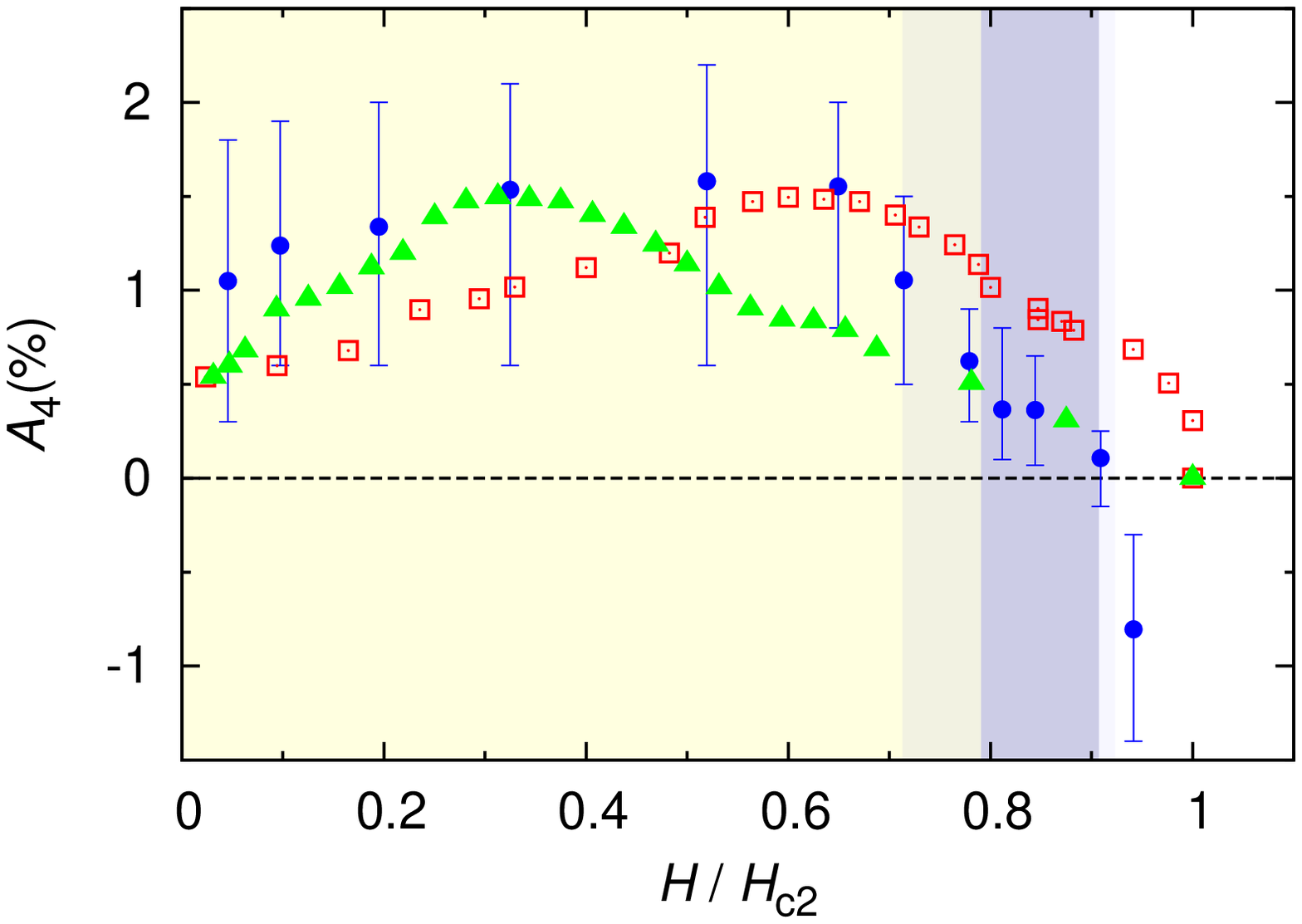} 
\caption{
Field dependence of $A_4$ calculated with $\mu=0$ (triangles) and $\mu=0.04$ (squares) by assuming a horizontal line node gap and a large in-plane $v_{\rm F}$ anisotropy ($b=0.5$). 
Circles are the experimental data at 0.1~K.
}
\label{calc}
\end{center}
\end{figure}

Figure \ref{calc} compares $A_4(H)$ obtained from calculations with (squares; $\mu=0.04$) and without (triangles; $\mu=0$) the Pauli-paramagnetic effect 
by assuming a horizontal line node gap and a large in-plane $v_{\rm F}$ anisotropy.
Inclusion of the Pauli-paramagnetic effect tends to shift a hump in $A_4(H)$ toward a higher field.
In the case of no Pauli-paramagnetic effect, $A_4(H)$ starts to decrease when the magnetic field exceeds $\sim 0.3 \Hc2$.
This behavior is in contrast to the experimental result at 0.1~K (circles in Fig.~\ref{calc}).
In addition, a rapid increase in $C_{\rm e}(H)$ with approaching $\Hc2$ cannot be reproduced if $\mu=0$ is adopted.
Thus, the Pauli-paramagnetic effect is favorable to explain the specific-heat behavior under a magnetic field.

\vspace{0.3in}
\hrule width 17.8cm
\vspace{0.1in}
\noindent
{\scriptsize
1)\ \  D. Shibata, H. Tanaka, S. Yonezawa, T. Nojima, and Y. Maeno,  Phys. Rev. B {\bf 91}, 104514 (2015).
}
\clearpage

\begin{thebibliography}{90}

\bibitem{Maeno1994Nature}
Y. Maeno, H. Hashimoto, K. Yoshida, S. Nishizaki, T. Fujita, J.~G. Bednorz, and
  F. Lichtenberg,  Nature (London) {\bf 372}, 532 (1994).
\bibitem{Ishida1998Nature}
K. Ishida, H. Mukuda, Y. Kitaoka, K. Asayama, Z.~Q. Mao, Y. Mori, and Y. Maeno,
   Nature (London) {\bf 396}, 658 (1998).
\bibitem{Duffy2000PRL}
J.~A. Duffy, S.~M. Hayden, Y. Maeno, Z. Mao, J. Kulda, and G.~J. McIntyre,
  Phys. Rev. Lett. {\bf 85}, 5412 (2000).
\bibitem{Ishida2008JPCS}
K. Ishida, H. Murakawa, H. Mukuda, Y. Kitaoka, Z.~Q. Mao, and Y. Maeno,  J.
  Phys. Chem. Solids {\bf 69}, 3108 (2008).
\bibitem{Ishida2015PRB}
K. Ishida, M. Manago, T. Yamanaka, H. Fukazawa, Z.~Q. Mao, Y. Maeno, and K.
  Miyake,  Phys. Rev. B {\bf 92}, 100502(R) (2015).
\bibitem{Mackenzie2003RMP}
A.~P. Mackenzie and Y. Maeno,  Rev. Mod. Phys. {\bf 75}, 657 (2003).
\bibitem{Maeno2012JPSJ}
Y. Maeno, S. Kittaka, T. Nomura, S. Yonezawa, and K. Ishida,  J. Phys. Soc.
  Jpn. {\bf 81}, 011009 (2012).
\bibitem{Luke1998Nature}
G.~M. Luke, Y. Fudamoto, K.~M. Kojima, M.~I. Larkin, J. Merrin, B. Nachumi,
  Y.~J. Uemura, Y. Maeno, Z.~Q. Mao, Y. Mori, H. Nakamura, and M. Sigrist,
  Nature (London) {\bf 394}, 558 (1998).
\bibitem{Xia2006PRL}
J. Xia, Y. Maeno, P.~T. Beyersdorf, M.~M. Fejer, and A. Kapitulnik,  Phys. Rev.
  Lett. {\bf 97}, 167002 (2006).
\bibitem{Bergemann2003AP}
C. Bergemann, A.~P. Mackenzie, S.~R. Julian, D. Forsythe, and E. Ohmichi,  Adv.
  Phys. {\bf 52}, 639 (2003).
\bibitem{Damascelli2000PRL}
A. Damascelli, D.~H. Lu, K.~M. Shen, N.~P. Armitage, F. Ronning, D.~L. Feng, C.
  Kim, Z.~X. Shen, T. Kimura, Y. Tokura, Z.~Q. Mao, and Y. Maeno,  Phys. Rev.
  Lett. {\bf 85}, 5194 (2000).
\bibitem{Mackenzie2017QM}
A.~P. Mackenzie, T. Scaffidi, C.~W. Hicks, and Y. Maeno,  npj Quantum Materials
  {\bf 2}, 40 (2017).
\bibitem{Yonezawa2013PRL}
S. Yonezawa, T. Kajikawa, and Y. Maeno,  Phys. Rev. Lett. {\bf 110}, 077003
  (2013).
\bibitem{Yonezawa2014JPSJ}
S. Yonezawa, T. Kajikawa, and Y. Maeno,  J. Phys. Soc. Jpn. {\bf 83}, 083706
  (2014).
\bibitem{Kittaka2014PRB}
S. Kittaka, A. Kasahara, T. Sakakibara, D. Shibata, S. Yonezawa, Y. Maeno, K.
  Tenya, and K. Machida,  Phys. Rev. B {\bf 90}, 220502(R) (2014).
\bibitem{Kikugawa2016PRB}
N. Kikugawa, T. Terashima, S. Uji, K. Sugii, Y. Maeno, D. Graf, R. Baumbach,
  and J. Brooks,  Phys. Rev. B {\bf 93}, 184513 (2016).
\bibitem{Nishizaki2000JPSJ}
S. Nishizaki, Y. Maeno, and Z. Mao,  J. Phys. Soc. Jpn. {\bf 69}, 572 (2000).
\bibitem{Suzuki2002PRL}
M. Suzuki, M.~A. Tanatar, N. Kikugawa, Z.~Q. Mao, Y. Maeno, and T. Ishiguro,
  Phys. Rev. Lett. {\bf 88}, 227004 (2002).
\bibitem{Bonalde2000PRL}
I. Bonalde, B.~D. Yanoff, M.~B. Salamon, D.~J.~van Harlingen, E.~M.~E. Chia,
  Z.~Q. Mao, and Y. Maeno,  Phys. Rev. Lett. {\bf 85}, 4775 (2000).
\bibitem{Ishida2000PRL}
K. Ishida, H. Mukuda, Y. Kitaoka, Z.~Q. Mao, Y. Mori, and Y. Maeno,  Phys. Rev.
  Lett. {\bf 84}, 5387 (2000).
\bibitem{Izawa2001PRL}
K. Izawa, H. Takahashi, H. Yamaguchi, Y. Matsuda, M. Suzuki, T. Sasaki, T.
  Fukase, Y. Yoshida, R. Settai, and Y. Onuki,  Phys. Rev. Lett. {\bf 86}, 2653
  (2001).
\bibitem{Tanatar2001PRL}
M.~A. Tanatar, M. Suzuki, S. Nagai, Z.~Q. Mao, Y. Maeno, and T. Ishiguro,
  Phys. Rev. Lett. {\bf 86}, 2649 (2001).
\bibitem{Deguchi2004PRL}
K. Deguchi, Z.~Q. Mao, H. Yaguchi, and Y. Maeno,  Phys. Rev. Lett. {\bf 92},
  047002 (2004).
\bibitem{Deguchi2004JPSJ}
K. Deguchi, Z.~Q. Mao, and Y. Maeno,  J. Phys. Soc. Jpn. {\bf 73}, 1313 (2004).
\bibitem{Hassenger2017PRX}
E. Hassinger, P. Bourgeois-Hope, H. Taniguchi, S.~R. de~Cotret, G.
  Grissonnanche, M.~S. Anwar, Y. Maeno, N. Doiron-Leyraud, and L. Taillefer,
  Phys. Rev. X {\bf 7}, 011032 (2017).
\bibitem{Sakakibara2016RPP}
T. Sakakibara, S. Kittaka, and K. Machida,  Rep. Prog. Phys. {\bf 79}, 094002
  (2016).
\bibitem{Matsuda2006JPCM}
Y. Matsuda, K. Izawa, and I. Vekhter,  J. Phys.:Condens. Matter {\bf 18}, R705
  (2006).
\bibitem{Vorontsov2006PRL}
A. Vorontsov and I. Vekhter,  Phys. Rev. Lett. {\bf 96}, 237001 (2006).
\bibitem{Hiragi2010JPSJ}
M. Hiragi, K.~M. Suzuki, M. Ichioka, and K. Machida,  J. Phys. Soc. Jpn. {\bf
  79}, 094709 (2010).
\bibitem{An2010PRL}
K. An, T. Sakakibara, R. Settai, Y. {$\bar{\mathrm{O}}$}nuki, M. Hiragi, M.
  Ichioka, and K. Machida,  Phys. Rev. Lett. {\bf 104}, 037002 (2010).
\bibitem{Kittaka2014JPSJ}
S. Kittaka, Y. Aoki, N. Kase, T. Sakakibara, T. Saito, H. Fukazawa, Y. Kohori,
  K. Kihou, C.~H. Lee, A. Iyo, H. Eisaki, K. Deguchi, N.~K. Sato, Y. Tsutsumi,
  and K. Machida,  J. Phys. Soc. Jpn. {\bf 83}, 013704 (2014).
\bibitem{Mao2000MRB}
Z.~Q. Mao, Y. Maeno, and H. Fukazawa,  Mat. Res. Bull. {\bf 35}, 1813 (2000).
\bibitem{Mackenzie1998PRL}
A.~P. Mackenzie, R.~K.~W. Haselwimmer, A.~W. Tyler, G.~G. Lonzarich, Y. Mori,
  S. Nishizaki, and Y. Maeno,  Phys. Rev. Lett. {\bf 80}, 161 (1998).
\bibitem{Ishida2001PRB}
K. Ishida, H. Mukuda, Y. Kitaoka, Z.~Q. Mao, H. Fukazawa, and Y. Maeno,  Phys.
  Rev. B {\bf 63}, 060507(R) (2001).
\bibitem{Murakawa2004PRL}
H. Murakawa, K. Ishida, K. Kitagawa, Z.~Q. Mao, and Y. Maeno,  Phys. Rev. Lett.
  {\bf 93}, 167004 (2004).
\bibitem{Kittaka2017c}
If the field-independent term of non-electronic contributions is adjusted by
  assuming that $C_{\rm e}/T$ of $\SRO$ in the normal state remains
  constant,\cite{Nishizaki2000JPSJ} the absolute value of $C_{\rm e}/T$ also
  becomes comparable to a previous report~\cite{Deguchi2004PRL}.
\bibitem{Mao2000PRL}
Z.~Q. Mao, Y. Maeno, S. NishiZaki, T. Akima, and T. Ishiguro,  Phys. Rev. Lett.
  {\bf 84}, 991 (2000).
\bibitem{Yaguchi2002PRB}
H. Yaguchi, T. Akima, Z. Mao, Y. Maeno, and T. Ishiguro,  Phys. Rev. B {\bf
  66}, 214514 (2002).
\bibitem{Volovik1993JETPL}
G.~E. Volovik,  JETP Lett. {\bf 58}, 469 (1993).
\bibitem{Bauquet2002PRL}
F. Bouquet, Y. Wang, I. Sheikin, T. Plackowski, A. Junod, S. Lee, and S.
  Tajima,  Phys. Rev. Lett. {\bf 89}, 257001 (2002).
\bibitem{Hardy2014JPSJ}
F. Hardy, R. Eder, M. Jackson, D. Aoki, C. Paulsen, T. Wolf, P. Burger, A.
  B\"ohmer, P. Schweiss, P. Adelmann, R.~A. Fisher, and C. Meingast,  J. Phys.
  Soc. Jpn. {\bf 83}, 014711 (2014).
\bibitem{Kittaka2018a}
From the present data, we cannot distinguish the oscillation pattern in
  details, \textit{e.g.}, $\cos(4\phi)$ and
  $1-2|\sin(2\phi)|$,\cite{Deguchi2004PRL} which contains useful information on
  the gap anisotropy.
\bibitem{KittakaSM}
(Supplemental Material) (I) Results of $C(\theta)$ measurements on
  Sr$_2$RuO$_4$, and (II) Pauli-paramagnetic effect on $A_4(H)$ are provided
  online.
\bibitem{Kittaka2017b}
In recent studies on URu$_2$Si$_2$~\cite{Kittaka2016JPSJ} and
  UPd$_2$Al$_3$,\cite{Shimizu2016PRL} evidence for the presence of horizontal
  line nodes has been successfully provided from $C(\theta)$ measurements,
  which have also been theoretically supported~\cite{Tsutsumi2016PRB}.
\bibitem{Vekhter2008PhysicaB}
I. Vekhter and A. Vorontsov,  Physica B {\bf 403}, 958 (2008).
\bibitem{Machida2017}
K. Machida, K. Irie, K. Suzuki, Y. Tsutsumi, and H. Ikeda, unpublished.
\bibitem{Steppke2017Science}
A. Steppke, L. Zhao, M.~E. Barber, T. Scaffidi, F. Jerzembeck, H. Rosner, A.~S.
  Gibbs, Y. Maeno, S.~H. Simon, A.~P. Mackenzie, and C.~W. Hicks,  Science {\bf
  355}, eaaf9398 (2017).
\bibitem{Kittaka2017d}
Recent experiment of scanning tunneling spectroscopy evaluated $v_{{\rm F}
  \parallel [110]}/v_{{\rm F} \parallel [100]} \approx 0.7$ for the $\beta$
  band~\cite{Wang2017NatPhys}.
\bibitem{Vekhter1999PRB}
I. Vekhter, P.~J. Hirschfeld, J.~P. Carbotte, and E.~J. Nicol,  Phys. Rev. B
  {\bf 59}, R9023 (1999).
\bibitem{Miranovic2005JPC}
P. Miranovi{$\acute{\mathrm{c}}$}, M. Ichioka, K. Machida, and N. Nakai,  J.
  Phys.: Condens. Matter {\bf 17}, 7971 (2005).
\bibitem{Amano2014PRB}
Y. Amano, M. Ishihara, M. Ichioka, N. Nakai, and K. Machida,  Phys. Rev. B {\bf
  90}, 144514 (2014).
\bibitem{Ichioka2007PRB}
M. Ichioka and K. Machida,  Phys. Rev. B {\bf 76}, 064502 (2007).
\bibitem{Machida2008PRB}
K. Machida and M. Ichioka,  Phys. Rev. B {\bf 77}, 184515 (2008).
\bibitem{Machida2017b}
$A_4(T,H)$ at finite $\mu$ has been evaluated from the data calculated with
  $\mu=0$ by using the universal scaling relation $[dN(E)/dE]_{E \simeq
  0}\propto 1-N(E=0)$~\cite{Machida2017}.
\bibitem{Suzuki2011JPSJ}
K.~M. Suzuki, Y. Tsutsumi, N. Nakai, M. Ichioka, and K. Machida,  J. Phys. Soc.
  Jpn. {\bf 80}, 123706 (2011).
\bibitem{Kittaka2009PRB}
S. Kittaka, T. Nakamura, Y. Aono, S. Yonezawa, K. Ishida, and Y. Maeno,  Phys.
  Rev. B {\bf 80}, 174514 (2009).
\bibitem{Kittaka2017e}
The $\alpha$ and $\beta$ bands have relatively small anisotropy in the
  amplitude of the in-plane $\Vec{v}_{\rm F}$~\cite{Steppke2017Science} but
  have large anisotropy in its orientation, reflecting the
  quasi-one-dimensional nature. This anisotropy also causes an oscillation of
  $C_{\rm e}(\phi=0^\circ)<C_{\rm e}(\phi=45^\circ)$ if horizontal line nodes
  exist~\cite{Machida2017}.
\bibitem{Zhitomirsky2001PRL}
M.~E. Zhitomirsky and T.~M. Rice,  Phys. Rev. Lett. {\bf 87}, 057001 (2001).
\bibitem{Kunkemoller2017PRL}
S. Kunkem$\ddot{\mathrm{o}}$ller, P. Steffens, P. Link, Y. Sidis, Z.~Q. Mao, Y.
  Maeno, and M. Braden,  Phys. Rev. Lett. {\bf 118}, 147002 (2017).
\bibitem{Kittaka2016JPSJ}
S. Kittaka, Y. Shimizu, T. Sakakibara, Y. Haga, E. Yamamoto, Y.
  {$\bar{\mathrm{O}}$}nuki, Y. Tsutsumi, T. Nomoto, H. Ikeda, and K. Machida,
  J. Phys. Soc. Jpn. {\bf 85}, 033704 (2016).
\bibitem{Shimizu2016PRL}
Y. Shimizu, S. Kittaka, T. Sakakibara, Y. Tsutsumi, T. Nomoto, H. Ikeda, K.
  Machida, Y. Homma, and D. Aoki,  Phys. Rev. Lett. {\bf 117}, 037001 (2016).
\bibitem{Tsutsumi2016PRB}
Y. Tsutsumi, T. Nomoto, H. Ikeda, and K. Machida,  Phys. Rev. B {\bf 94},
  224503 (2016).
\bibitem{Wang2017NatPhys}
Z. Wang, D. Walkup, P. Derry, T. Scaffidi, M. Rak, S. Vig, A. Kogar, I.
  Zeljkovic, A. Husain, L.~H. Santos, Y. Wang, A. Damascelli, Y. Maeno, P.
  Abbamonte, E. Fradkin, and V. Madhavan,  Nat. Phys. {\bf 13}, 799 (2017).
\end{thebibliography}
\end{document}